\documentclass[11pt, oneside]{article}   	
\usepackage{geometry}                		
\usepackage{graphicx}				

\usepackage{amsmath,amssymb,amsthm,mathtools}
\usepackage[hidelinks]{hyperref}
\newcommand{\T}{\mathrm T}
\newcommand{\Tr}{\operatorname{Tr}}

\newcommand{\cL}{\mathcal L}
\newcommand{\cV}{\mathcal V}

\newcommand{\x}{{\bf x}}

\title{ Induction instead of Feynman diagrams, L-functionals instead of  Hilbert space}

\usepackage{booktabs,array}
\newcommand{\STr}{\operatorname{STr}}
\newcommand{\Real}{\operatorname{Re}}
\usepackage{listings}
\lstdefinestyle{properpython}{language=Python,
 basicstyle=\ttfamily\footnotesize,columns=fullflexible,
 keepspaces=true,showstringspaces=false,breaklines=true,
 frame=single,numbers=left,numberstyle=\tiny,numbersep=6pt,
 xleftmargin=3em,framexleftmargin=2.5em,tabsize=4}
\begin{document}
\author {  Albert Schwarz\\ Department of Mathematics\\ 
University of 
California \\ Davis, CA 95616, USA,\\ schwarz @math.ucdavis.edu}

\maketitle
\begin {abstract} 
{The main result of this paper is a construction of an inductive process that can replace Feynman diagrams.
Combining this result with the notion of the L-functional, one can give an exposition of QED that does not suffer from
 infrared divergences.}
 \end{abstract}
  
 \section{Introduction}
 Let us consider Green functions or generalized Green functions (GGreen functions, see the definition in Section 3). They can be represented as a sum of Feynman diagrams. A connected Green function can be defined as a sum of connected diagrams.  We will consider amputated connected Green functions (sums of connected diagrams with deleted external edges). A proper Green function is defined as a partial sum of the sum representing the amputated connected Green function that consists
 of 1 PI diagrams. (Recall that a 1PI diagram is a diagram that stays connected when we delete any  of its internal edges.) Knowing proper Green functions, we can restore both connected and full Green functions. One can say that proper Green functions determine the effective action; this means that connected Green functions are sums of tree diagrams where vertices are proper Green functions.
 
 Let us denote by $\Gamma_{L, m}$ a sum of all amputated 1PI diagrams having $L$ linearly independent loops and $m$ internal edges. Our induction is based on the following formula
 \begin{equation}
 \Gamma_{L,m}=
 \frac1{2m}\sum_{k=1}^{m}
 \ \sum_{\substack{L_1+\cdots+L_k=L-1\\
                    m_1+\cdots+m_k=m-k}}
 Tr \!\left[
 P\Gamma_{L_1,m_1}^{(2)}\cdots
 P\Gamma_{L_k,m_k}^{(2)}\right] .
 \label{eq:rec}
\end{equation}
Here $L,m>0$,all inner indices are nonnegative. Every lower block has loop number
strictly smaller than $L$, so this is an induction in proper functions.

In the above formula $P=P^{ab}$ stands for the propagator, $F^{(2)}$ stands for the second derivative. In other words,
 
 $Tr \!\left[
 P\Gamma_{L_1,m_1}^{(2)}\cdots
 P\Gamma_{L_k,m_k}^{(2)}\right] =
 P^{ab}\partial_{bc}\Gamma_{L_1,m_1}\cdots
 P^{rs}\partial_{sa}\Gamma_{L_k,m_k}.$
 
 The same formula is valid for GGreen functions. If there are fermions in the theory, one should replace the trace by supertrace.
 
 There exists a proof of the formula (1) that is based on the definition of proper Green (or GGreen) functions in terms of diagrams. In this proof, we delete one of the internal edges of the diagram and express the contribution of the diagram in terms of the remainder. The remainder is not necessarily a 1PI diagram, but it can be represented as a chain of 1PI blocks. This remark leads to the formula (1). The coefficient $1/2m$ in this formula comes from the consideration of automorphism groups of the diagrams we consider.

 In Section 2 we will give more details of this proof and another proof that allows us to avoid Feynman diagrams.
 
 In  Section 3 we will review generalized Green functions (GGreen functions) and their relation to inclusive cross sections following mostly \cite {B}. We will explain how GGreen functions appear naturally in the formalism of L-functionals.
 
 Finally, Appendix  A (written by ChatGPT) contains an application of our methods to QED. It is written as an exposition of QED that is very different from existing QFT textbooks. The conventional notion of the scattering matrix is not applicable in QED because there exist no processes that involve only a finite number of particles (due to the existence of soft photons). Nevertheless, it is used in textbooks. A state of an electron dressed by a photon cloud does not belong to Fock space; this is a source of infrared divergences. These divergences disappear if we work in terms of L-functionals and GGreen functions.
 
 We use induction to analyze inclusive cross sections in QED for textbook problems in leading order (LO) and next-to-leading order (NLO). 
 
 Our calculations are based on the expression of inclusive cross sections in terms of GGreen functions
 \cite {NP}, \cite {B}. This expression was given for theories with massive particles, but it is well known that the inclusive cross sections in QED are limits of the inclusive cross sections in the theory with a massive photon. 
 
 Appendix B  ( also written by ChatGPT) contains code for calculations with the formula (1).
 
 Our results do not use locality.  They could be useful in statistical physics. To apply them to local theories, we should introduce UV cutoff.
  
  The induction based on the formula (1) allows us to represent proper Green and GGreen functions as integrals; similar integrals we obtain using Feynman diagrams. For large $L$, the number of 1PI diagrams grows faster than $L!$. From the other side, one can hope that the computational complexity of induction grows polynomially with $L$.
 In any case, even for small $L$, the inductive calculations are slightly shorter than the calculations with Feynman diagrams.  I am quoting the opinion of ChatGPT: 
 
"For fixed external multiplicity, I expect little computational advantage at one or two loops, but potentially very large savings at high loop order. The important qualification is that we can presently prove a polynomial bound for the induction’s functional bookkeeping, rather than for all elementary operations needed to evaluate a QED momentum integrand."

 \section {The proof of the induction formula}.
 
 Let us start with more details of the proof sketched in the Introduction.
 
 By definition, an edge of a connected graph is a bridge if removing this edge gives a disconnected graph.  In a  1PI graph, there are no bridges. However, deleting an edge of 1PI graph can lead to a graph containing bridges. This new graph, like any graph, can be considered as a tree where vertices are 1PI blocks. In our situation, there exists a chain that connects the vertices of the deleted edge. (Obviously, such a chain exists inside the new graph. It should coincide with this graph because every edge outside the chain is a bridge of the original graph.)
 
 The formula  \begin{equation}\label {D}
 P^{ab}\partial_{bc}\Gamma_{L_1,m_1}\cdots
 P^{rs}\partial_{sa}\Gamma_{L_k,m_k}
 \end{equation}
  expresses the contribution of 1PI graphs in terms of simpler graphs. However, this formula is written for a distinguished edge. To get the formula (1), we should take into account that we have $m$ edges. Every edge has two orientations. A distinguished edge is oriented; therefore, to avoid overcounting, we should divide by $2m$.
  
  To give a more accurate proof, we should recall that the contribution of every diagram $D$  contains a factor $1/|Aut(D)|$, where $|Aut(D)|$ stands for the number of elements in the group $Aut(D)$ of automorphisms of the diagram. If we have a distinguished edge $\mu$ we should divide the contribution by $|Aut (D,\mu)|$ where $Aut (D,\mu)$ consists of the elements of $Aut(D)$ that transform an oriented edge $\mu$ into itself.  One can say that $Aut (D,\mu)$ is a stabilizer of the group $Aut(D)$ acting on the set of oriented edges. It follows that

 $$\frac {1}{|Aut (D,\mu)|}=\frac {2m}{|Aut(D)|}.
 $$
 This means that (\ref {D})  implies (1).
 
 Let us give another proof of the induction formula that uses a different definition of $\Gamma_{L,m}$. For simplicity, in what follows  we will work with a finite number of variables; however, the generalization to an infinite number of variables is not difficult.
 
 Introduce the operator
\begin{equation}
 \Delta_P=\frac12 P^{AB}
 \frac{\delta^2}{\partial\chi^A\partial\chi^B},
 \qquad P^{AB}=P^{BA},
 \label{eq:Delta}
\end{equation}
Here $P$ is a non-degenerate matrix. Its inverse is denoted by $Q.$

The diagram technique can be represented by the formula

\begin{equation}
 G=
 \left.e^{\Delta_P}
 \exp\!\left({\cV[\chi]}\right)
 \right|_{\chi=0}.
 \label{G}
\end{equation}
where $\cV$ is a polynomial with respect to $\chi$.

It is easy to check that $G$ given by this formula is represented as a sum of diagrams with the propagator $P^{AB}$ and vertices coming from $\cV$.

One can obtain (\ref {G}) without reference to diagrams as the formula that governs the perturbation theory for the integral of the expression $\exp(-\frac 1 2 \chi Q\chi +\cV(\chi))$ where  $\cV$ is considered as perturbation.

Let us assume that $\chi$ in (\ref{G}) rus over finite-dimensional space $\mathbb {R}^N.$
Since
\[
 \Delta_P e^{ip_A\chi^A}
   =-\frac12p_A P^{AB}p_B\,e^{ip_A\chi^A},
\]
the operator $e^{\Delta_P}$ becomes multiplication by a Gaussian
after Fourier transformation.

Taking the inverse Fourier transform gives
\begin{align*}
 (e^{\Delta_P}f)(\chi)
 &=\frac1{(2\pi)^N}\int d^N p\,
    e^{ip_A\chi^A-\frac12p_A P^{AB}p_B}\widehat f(p)\\
 &=\int d^N x\,f(x)\,
   \frac1{(2\pi)^N}\int d^N p\,
    e^{ip_A(\chi^A-x^A)-\frac12p_A P^{AB}p_B}.
\end{align*}
Calculating the Gaussian integral, we obtain
\[
  (e^{\Delta_P}f)(\chi)
 =\frac1{\mathcal N_P}\int_{\mathbb R^N}d^N x\,
   e^{-\frac12(x-\chi)^A Q_{AB}(x-\chi)^B}f(x).
\]
where
\[
 \mathcal N_P=(2\pi)^{N/2}\sqrt{\det P}
\]

Taking $f=e^{\cV}$ and then $\chi=0$ we obtain

$$G=\frac {1}{\mathcal N_P}\int dx e^{-\frac12x^A Q_{AB}x^B+\cV(x)}$$.

Our calculations are valid only under some conditions imposed on  $P$ and $\cV$.  However, a slight modification of our arguments shows that the above formula is true for every polynomial $\cV$ in the framework of perturbation theory with respect to $\cV.$

To prove the induction formula, we  introduce new parameters in $G$:
\begin{equation}
 G_{\tau,z}[J]=
 \left.e^{z\tau\Delta_P}
 \exp\!\left(\frac{\cV[\chi]+J_A\chi^A}{z}\right)
 \right|_{\chi=0}.
 \label{eq:Gtauz}
\end{equation}
All expressions are understood as formal perturbative series.
 At $\tau=z=1$, we obtain (\ref{G}).

Again, we assume that our functions depend on a finite number of variables $N$ ; the number $N$ does not enter the final formula for induction.
For an arbitrary formal function $f(\chi)$, define
\[
 \langle f\rangle_{\tau,z}
 =\left.e^{z\tau\Delta_P}f[\chi]\right|_{\chi=0}.
\]
The elementary Gaussian identities are
\begin{align}
 \langle\chi^A f\rangle_{\tau,z}
 &=z\tau P^{AB}\langle\partial_Bf\rangle_{\tau,z},
 \label{eq:WickIBP}\\
 \partial_\tau\langle f\rangle_{\tau,z}
 &=\frac z2P^{AB}\langle\partial_A\partial_B f\rangle_{\tau,z}
 =\frac{1}{2z\tau^2}
   \langle\chi^A Q_{AB}\chi^B f\rangle_{\tau,z}
   -\frac{N}{2\tau}\langle f\rangle_{\tau,z}.
 \label{eq:Wickflow}
\end{align}
The first follows by commuting multiplication by $\chi^A$ through
$e^{z\tau\Delta_P}$. Applying it twice proves the second equality
in \eqref{eq:Wickflow}. 

Set
\[
 D^A=z\frac{\partial}{\partial J_A},\qquad
 \cV_A[\chi]=\frac{\partial\cV[\chi]}{\partial\chi^A}.
\]
Substitution of $f=\exp((\cV+J\chi)/z)$ into
\eqref{eq:WickIBP} and \eqref{eq:Wickflow} yields two linear
equations for $G_{\tau,z}$:
\begin{align}
 \left(\frac1\tau Q_{AB}D^B-J_A-\cV_A[D]\right)
 G_{\tau,z}[J]&=0,
 \label{eq:GSD}\\
 \partial_\tau G_{\tau,z}[J]
 &=\left(\frac{z}{2\tau^2}Q_{AB}
       \frac{\partial^2}{\partial J_A\partial J_B}
       -\frac{N}{2\tau}\right)G_{\tau,z}[J].
 \label{eq:Gflow}
\end{align}
The apparent negative powers of $\tau$ cancel for the functional
\eqref{eq:Gtauz}. For example, when $\cV=0$,
\[
 G_{\tau,z}[J]=\exp\!\left(\frac{\tau}{2z}JPJ\right),
\]
and both equations are immediate. 
Use the scaled logarithm
\begin{equation}
 W_{\tau,z}[J]=z\log G_{\tau,z}[J],\qquad
 \varphi^A=\frac{\partial W_{\tau,z}}{\partial J_A},\qquad
 C^{AB}=\frac{\partial^2 W_{\tau,z}}{\partial J_A\partial J_B}.
 \label{eq:Wdefinition}
\end{equation}
At $z=1$, $W_{\tau,1}$ is the logarithm itself. The factor $z$
is convenient because a connected graph with $L$ loops then has
weight $z^L$.
The identities
\[
 G^{-1}D^AG=\varphi^A,\qquad
 G^{-1}D^AD^BG=\varphi^A\varphi^B+z C^{AB}
\]
turn \eqref{eq:Gflow} into
\begin{equation}
 \partial_\tau W_{\tau,z}
 =\frac1{2\tau^2}\varphi Q\varphi
  +\frac{z}{2\tau^2}\Tr(QC)-\frac{zN}{2\tau}.
 \label{eq:Wflow}
\end{equation}
Similarly, conjugating \eqref{eq:GSD} by $e^{W/z}$ gives
\begin{equation}
 \frac1\tau Q_{AB}\varphi^B
 =J_A+\cV_A\!\left[\varphi+
          z\frac{\partial}{\partial J}\right]1.
 \label{eq:WSD}
\end{equation}
Here the last term is an operator applied to the constant $1$:
each occurrence of $\chi^A$ in $\cV_A$ is replaced by
$\varphi^A+z\partial/\partial J_A$, and derivatives act on everything
to their right. 

Now we would like to apply the Legendre transform to $W_{\tau,z}[J].$  Recall that the Legendre transform $\Gamma(\varphi)$ of the function $W(J)$ is defined as the value of expression  $W(J)-
J\varphi$ at the stationary point. The Legendre transform is involutive: taking the Legendre transform of $\Gamma(\varphi)$ we come back to $W(J)$. 

Our Legendre transform is formal ( defined in the framework of perturbation theory). We can use the formal Legendre transform because $W_{\tau,z}[J]=\frac {\tau}{ 2} JPJ$ for $\cV=0$ is a non-degenerate quadratic form.

By definition
\begin{equation}
 \Gamma^{\mathrm{full}}_{\tau,z}[\varphi]
 =W_{\tau,z}[J]-J_A\varphi^A,
 \qquad \varphi^A=\frac {\partial}{\partial J^A}W_{\tau,z}[J].
 \label{eq:Legendre}
\end{equation}
For $\cV=0$ we have $\Gamma^{\mathrm{full}}_{\tau,z}[\varphi]=-\frac1{2\tau}\varphi Q\varphi$.
Remove the known free quadratic term by defining
\begin{equation}
 \Gamma^{\mathrm{full}}_{\tau,z}[\varphi]
 =-\frac1{2\tau}\varphi Q\varphi
   +\Gamma_{\tau,z}[\varphi].
 \label{eq:interactionGamma}
\end{equation}
Then
\begin{equation}
 J=\frac1\tau Q\varphi-\Gamma^{(1)},\qquad
 C=\left(\frac1\tau Q-\Gamma^{(2)}\right)^{-1}.
 \label{eq:sourceHessian}
\end{equation}
Now \eqref{eq:WSD} becomes an equation entirely in
$\varphi$:
\begin{equation}
 \Gamma^{(1)}_A[\varphi]
 =\cV_A\!\left[\varphi+z C\frac{\partial}{\partial\varphi}\right]1,
 \qquad C=\left(\frac1\tau Q-\Gamma^{(2)}\right)^{-1}.
 \label{eq:GammaSD}
\end{equation}
(We use the notation $F^{(1)}$ for first derivative and $F^{(2)}$ for  second derivative.)
In components, the operator replacing $\chi^A$ is
$\varphi^A+z C^{AB}\partial/\partial\varphi^B$;
the same ordering convention as in \eqref{eq:WSD} applies.

The equation that gives the induction is simpler. Substituting
\eqref{eq:interactionGamma} into \eqref{eq:Wflow}, the derivatives
of the free quadratic terms cancel. Hence
\begin{align}
 \partial_\tau\Gamma_{\tau,z}
 &=\frac{z}{2\tau^2}\bigl(\Tr(QC)-\tau N\bigr)\nonumber\\
 &=\frac{z}{2\tau}\Tr\!\left[
       (1-\tau P\Gamma_{\tau,z}^{(2)})^{-1}-1\right].
 \label{eq:Gammaflowintermediate}
\end{align}
We used \eqref{eq:sourceHessian} and cyclicity of the trace.
The two terms containing $N=\Tr 1$ have canceled before any
infinite-dimensional limit is taken. Using
$(1-A)^{-1}-1=A(1-A)^{-1}$, we obtain
\begin{equation}
 \partial_\tau\Gamma_{\tau,z}[\varphi]
 =\frac z2\Tr\!\left[
 P\Gamma_{\tau,z}^{(2)}[\varphi]
 (1-\tau P\Gamma_{\tau,z}^{(2)}[\varphi])^{-1}
 \right].
 \label{eq:Gammaflow}
\end{equation}
The inverse in this equation is understood as a formal geometric series.

The initial condition is
\begin{equation}
 \Gamma_{0,z}[\varphi]=\cV[\varphi].
 \label{eq:Gammainitial}
\end{equation}
It is important to take this limit at fixed $\varphi$, rather than
at fixed $J$. Here is a direct verification. Translating the formal
Gaussian variable by $\varphi$ in \eqref{eq:Gtauz} and using
\eqref{eq:sourceHessian} gives
\begin{equation}
 e^{\Gamma_{\tau,z}[\varphi]/z}
 =\left.e^{z\tau\Delta_P}
 \exp\!\left(\frac{\cV[\varphi+\chi]
              -\Gamma^{(1)}_{\tau,z}[\varphi]\chi}{z}\right)
 \right|_{\chi=0}.
 \label{eq:Gammaimplicit}
\end{equation}
At $\tau=0$ the right-hand side is $e^{\cV[\varphi]/z}$.
Taking the formal logarithm proves \eqref{eq:Gammainitial}.
This identity also determines a regular series in $\tau$:
its term of degree zero is $\cV$, and its term of degree $r>0$
uses only coefficients of $\Gamma^{(1)}$ of lower degree.

\subsection{Induction relation}

Expand the solution of \eqref{eq:Gammaflow}--\eqref{eq:Gammainitial}
as
\begin{equation}
 \Gamma_{\tau,z}[\varphi]
 =\sum_{L,m\geq0}z^L\tau^m\Gamma_{L,m}[\varphi].
 \label{eq:Gammaexpansion}
\end{equation}
The initial condition and the factor $z$ in \eqref{eq:Gammaflow}
give
\begin{equation}
 \Gamma_{0,0}=\cV,\qquad
 \Gamma_{0,m}=0\quad(m>0),\qquad
 \Gamma_{L,0}=0\quad(L>0).
 \label{eq:initialcoefficients}
\end{equation}
They also show recursively that no negative powers of $z$ occur.
Expanding the inverse in \eqref{eq:Gammaflow} yields
\begin{equation}
 \partial_\tau\Gamma_{\tau,z}
 =\frac z2\sum_{k\geq1}\tau^{k-1}
 \Tr\!\left[(P\Gamma_{\tau,z}^{(2)})^k\right].
 \label{eq:Gammaflowseries}
\end{equation}
The coefficient of $z^L\tau^{m-1}$ on the left is
$m\Gamma_{L,m}$. A product of $k$ Hessians on the right contributes
to this coefficient precisely when
\[
 L_1+\cdots+L_k=L-1,\qquad
 m_1+\cdots+m_k=m-k.
\]
In particular $1\leq k\leq m$. Thus
\begin{equation}
 m\Gamma_{L,m}
 =\frac12\sum_{k=1}^{m}
 \sum_{\substack{L_1+\cdots+L_k=L-1\\
                  m_1+\cdots+m_k=m-k}}
 \Tr\!\left[
 P\Gamma_{L_1,m_1}^{(2)}\cdots
 P\Gamma_{L_k,m_k}^{(2)}\right].
 \label{eq:coefficientidentity}
\end{equation}
Dividing by $m$ proves \eqref{eq:rec}. Every $L_i\leq L-1$, so
this determines each loop order from the preceding loop orders.
For fixed $L,m$, all sums in \eqref{eq:coefficientidentity} are
finite.

So far, we have not used any diagrams. Now we will explain the relation of $\Gamma_{L,m}$ to 1PI diagrams.
 Expanding
\eqref{eq:Gtauz}, each propagator contributes $z\tau P$,
each interaction vertex contributes $\cV/z$, and each source
insertion contributes $J/z$. A connected graph containing an
interaction vertex, with $m$ internal edges, $v$ interaction vertices,
and $n$ external source edges
therefore has weight
\[
 (z\tau)^{m+n}z^{-v-n}
 =z^{L-1}\tau^{m+n},\qquad L=m-v+1.
\]
The logarithm retains connected graphs, and multiplication by $z$
changes this to $z^L$. The Legendre transform removes the connections
between proper parts, and its external arguments are amputated.
Indeed \eqref{eq:sourceHessian} is equivalent to
\[
 \varphi=\tau PJ+\tau P\Gamma^{(1)}_{\tau,z}[\varphi].
\]
Iteration expresses connected functions as trees made from the
proper vertices. Conversely, deleting the bridges of any connected
graph decomposes it uniquely into its maximal 1PI parts, joined as
a tree. These two inverse constructions identify the interaction
Legendre transform with the sum of amputated 1PI graphs. Each
amputation removes a factor $\tau P$ at an external argument;
therefore their remaining weights are exactly $z^L\tau^m$.
This proves the claimed identification of $\Gamma_{L,m}$.

As a check, at one loop the initial conditions force every block
in \eqref{eq:coefficientidentity} to be $\cV$. Consequently
\begin{equation}
 \Gamma_{1,m}=\frac1{2m}\Tr[(P\cV^{(2)})^m],\qquad
 \sum_{m\geq1}\tau^m\Gamma_{1,m}
 =-\frac12\Tr\log(1-\tau P\cV^{(2)}).
 \label{eq:oneloopcheck}
\end{equation}

Although we used $P^{-1}$ in the intermediate equations, every
coefficient in \eqref{eq:coefficientidentity} is polynomial in $P$.
The identity therefore extends algebraically to degenerate
covariances. In the field-theoretic setting the capital indices
again include space-time integration.
Keeping the momentum that is integrated in each trace as an
independent variable gives the corresponding inductive
construction of integrands. 

For fermions one uses Grassmann fields and sources, with the
Hessians interpreted as graded bilinear forms. With consistent
left and right Grassmann derivatives, the same proof uses the
supertrace
\[
 \operatorname{STr}M=\sum_A(-1)^{|A|}M^A{}_A,
\]
where $|A|=0$ for a bosonic component and $|A|=1$ for a fermionic
component. Replacing $\Tr$ by $\operatorname{STr}$ in
\eqref{eq:Gammaflow} and \eqref{eq:coefficientidentity} gives
precisely the fermionic version of \eqref{eq:rec}.

Our considerations can be applied also in BRST formalism.

\section{Green and GGreen functions. Inclusive cross sections. L-functionals.}
In this section we will remind the definition of generalized Green functions (GGreen functions), corresponding diagram techniques and relation to inclusive cross sections. We will  show that this notion appears naturally when states are represented by L-functionals instead of vectors in Hilbert spaces
 (for more detail see\cite {B}).
 
 In the algebraic approach, a state of the quantum system can be represented as a positive linear functional $\omega$ on the algebra of observables $\cal A$ (on a unital associative algebra with involution=$*$-algebra). A vector $\Phi$ or a density matrix $K$ in a representation space $E$ of $\cal A$ specify states by the formulas
  $\omega_{\Phi}(A)=\langle \Phi, A\Phi\rangle$, $\omega_K=Tr AK$. Conversely, every state can be represented by a vector in some representation of $\cal A$ (GNS construction).
 
 Every element $B\in\cal A$ specifies two operators on linear functionals on $\cal A$: operator $B^+$ transforms the fuctional $\omega(X)$ into the functional $\omega(XB)$,
 operator $B^-$ transforms the fuctional $\omega(X)$ into the functional $\omega(B^*X).$
 
 If we are interested in QFT, we need an action of spatial and temporal translations on $\cal A$. In other words,  we need a homomorphism of the $(d+1)$-dimensional abelian Lie group $\cal T$ of translations into the automorphism group of the $*$-algebra $\cal A$. The element of $\cal T$ will be written as $(t,\bf x)$ where ${\bf x}\in \mathbb{R}^d$ specifies a spatial translation, $t\in \mathbb R$ specifies a temporal translation. We use the notation $A({\bf x},t)$ for the translated $A$. The action of $\cal T$ on $\cal A$ induces the action of $\cal T$ on linear functionals. We will consider a state $\omega$ that is invariant with respect to spatial and temporal translations. To define the corresponding GGreen functions, we consider a chronological product $\T(MN)$ (times decreasing) where $M$ is a product of expressions of the form $A_i^+( t_i, {\bf x}_i)$ and $N$ is a product of expressions of the form $B_j^-(t_j,{\bf x}_j)$. The GGreen function corresponding to the state $\omega$  is defined as the value of the linear functional $\T(MN)\omega$ at the unit element $1$ of the algebra $\cal A$. If $N=1$ we obtain a Green function. In QFT we assume that $\omega$ is the ground state, but in non-equilibrium statistical physics (where GGreen functions were introduced by Keldysh) $\omega$ can stand for any translation-invariant stationary state.
 
 One can make the Fourier transform with respect to spatial variables and  the inverse Fourier transform with respect to time variables to obtain GGreen functions in the energy -momentum representation. The two-point Green function in this representation has the form $G(E_1,{\bf p}_1)\delta ({\bf p}_1-{\bf p}_2)\delta(E_1-E_2).$
 
 It is convenient to study QFT  using the representation of $\cal A$ in a Hilbert space $\cal E$ containing a cyclic vector $\Omega$ that obeys $\omega(A)=\langle \Omega, A\Omega\rangle$ (GNS construction). The group $\cal T$ of spatial and temporal translations acts in $\cal E$; infinitesimal translations are identified with the momentum $\bf P$ and Hamiltonian $H$. We assume that $H$ is nonnegative (this means that $\omega$ is a ground state). By definition, a particle is a vector function $\Phi(\bf p)$ obeying
 ${\bf P}\Phi({\bf p})={\bf p}\Phi({\bf p}), H\Phi({\bf p})=\epsilon ({\bf p})\Phi({\bf p})$. (More precisely, $\Phi({\bf p})$ is a generalized vector function. This means that only an integral $\Phi(f)$ of a product of $\Phi({\bf p})$ by some test function $f(\bf p)$ makes sense as an element of $\cal E$.) One says that $\epsilon (\bf p)$ is the dispersion law. ( Notice that we did not assume Lorentz invariance or locality.) It is important to emphasize that an electron in QED is not a particle in this sense.
 
 We define the one-particle subspace ${\cal E}_1$ as a subspace spanned by all vectors $\Phi(f)$.

 Particles correspond to the poles of the functions $G(E,{\bf p})$ with respect to $E$. We will assume that for every $\bf p$ this function has a unique pole $\epsilon (\bf p)$. 
 
 It is convenient to work in the formalism of L-functionals. This formalism can be applied if the phase space  of our theory is a vector space.

Let $\omega$ be a state represented as a linear functional on the Weyl or Clifford algebra  (an algebra generated by elements obeying CCR or CAR).  Its $L$-functional is defined by the formula
\begin{equation}
 \cL_\omega[f]=\omega\!\left(\exp\!\left(i\psi(f)\right)\right),\qquad
  \label{eq:L}
\end{equation}
where $\psi(f)$ is a Hermitian linear expression in terms of generators. It is convenient to write $\psi(f)$ as $\int dx f(\x)\psi(\x)$ and interpret $\psi(\x)$ as a field. (Both $f(\x)$ and $\psi(\x)$ can depend on some discrete arguments.)

The alternative convention with $i/\hbar$ in the exponential is obtained
by replacing $f$ by $f/\hbar$. We keep $\hbar$ explicit in the dynamics.
When a state is represented by a density matrix $\rho$, we have $\omega(B)=\Tr(\rho B)$; no density-matrix
realization is needed for the definitions.

Throughout, the Hamiltonian
        $\hat H$ is a formal expression in terms of $\psi(\x)$; we assume  that it induces a well-defined evolution on $L$-functionals.  The operator governing the evolution of $ L$- functionals will be denoted by $H$; it has the form $-\frac{i}{\hbar}(\hat H^+-\hat H^-)$ ( the definition of $\hat H^+$ and $\hat H^-$ on the space of linear functionals on the algebra is given above; we identify $L$-functionals with such linear functionals).

We assume that the equation of motion for $L$-functionals has the form
\begin{equation}
 \partial_t\cL_t=H\cL_t,\qquad
 H=-\frac{i}{\hbar}(\hat H^+-\hat H^-)=H_0+V.
 \label{eq:generator}
\end{equation}
where $H_0$ corresponds to the free Hamiltonian $\hat H_0$.

Put $H_a(t)=H_0+h(at)V$, with $h(t)=1$ for $t>T_1$ and $h(t)=0$ for $t<T_0$, where $T_0<T_1<0$. Let us consider adiabatic evolution
\begin{equation}
 \partial_t U_a(t,s)=H_a(t)U_a(t,s),\qquad U_a(s,s)=1.
\end{equation}
  We expect that the ground state of our theory can be obtained from the ground state $\omega_0=\cL_0$ of the free Hamiltonian by means of adiabatic dressing:
\begin{equation}
 \cL_\Omega=\lim_{a\downarrow0}U_a(0,-\infty)\cL_0,\qquad
 H\cL_\Omega=0.
 \label{eq:dressing}
\end{equation}


We derive the functional $G(J_+,J_-)$  generating GGreen functions directly from evolution on the
space of $L$-functionals. We then take the formal Legendre transform of its logarithm to get  a functional $\Gamma$ generating proper GGreen functions.  We introduce additional parameters to obtain the decomposition $\Gamma=\sum \Gamma_{L,m}.$ This allows us to 
prove the formula (1). 

Write $X=(t,\x)$. Let $J_+(t,\x),J_-(t,\x)$ have compact
time support and define
\begin{equation}
 H_{a,J}(t)=H_a(t)+\sum_{\sigma=\pm}\int d \x\,
 J_\sigma(t,\x)\psi^\sigma(\x).
 \label{eq:sources}
\end{equation}
Denote the corresponding evolution operator by $U_{a,J}$. Our sources are formal generating
variables. Physical real sources $j_\pm$ in exponentials with factors
$\pm i/\hbar$ correspond to $J_+=ij_+/\hbar$ and $J_-=-ij_-/\hbar$.

The generating functional for GGreen functions is given by the formula
\begin{equation}
 {G[J_+,J_-]=\lim_{a\downarrow0}
 (U_{a,J}(+\infty,-\infty)\cL_0)(1).}
 \label{eq:G}
\end{equation}

This follows immediately from the following statement. Let us consider the evolution operator $U(t,t_0)$ corresponding to the operator $A(t)$ and the evolution operator $\tilde U(t,t_0)$ corresponding to the operator $A(t)+a(t)$. Then

$$\tilde U(t,t_0)= U(t,t_0) \T\exp \int_{t_0}^t ds U (s,t_0)^{-1} a(s)U(s,t_0).$$

Using adiabatic dressing, we can derive a diagram technique for GGreen functions. The propagator of this technique  is a matrix consisting of free Green's functions:
\begin{align}
 P_{++}(X,Y)&=\omega_0(\T\psi_0(X)\psi_0(Y)),&
 P_{--}(X,Y)&=\omega_0(\T^{opp}\psi_0(X)\psi_0(Y)),\nonumber\\
 P_{+-}(X,Y)&=\omega_0(\psi_0(Y)\psi_0(X)),&
 P_{-+}(X,Y)&=\omega_0(\psi_0(X)\psi_0(Y)).
 \label{eq:P}
\end{align}
Here $\T^{opp}$ stands for antichronological product (times increasing).

The interaction $\hat V$ determines vertices of two types ($+$ and $-$ vertices).
We do not need a detailed description of vertices.
The interaction $\hat V$ determines vertices of two types ($+$ and $-$ vertices).

To express amplitudes or inclusive cross sections in terms of  Green or GGreen functions we should use fields $A(t,\bf x)$  that have the follwing property: the scalar product $\langle
\Phi({\bf p}),A\Omega\rangle$ does not vanish (hence  $A\Omega$ has a non-vanishing projection on one-particle space).

We say that the amputated Green or GGreen function in energy-momentum representation is on shell
if  the amputated edges have energy-momentum $(\epsilon ({\bf p}),{\bf p})$ where $\epsilon ({\bf p})$
stands for one-particle energy.

The scattering amplitudes can be expressed in terms of Green functions on shell (LSZ formula).
Similarly, inclusive cross sections can be expressed in terms of GGreen functions on shell (\cite {NP},\cite {B}). It is important to notice that the expression for the scattering amplitude in LSZ formula contains a factor $Z^{\frac 1 2}$ that depends on the choice of fields entering the definition of Green function.
The expression of the inclusive cross section in terms of the GGreen function does not contain such a factor.

All these statements were proven for scattering of usual particles under the assumption that the theory has strong cluster property. ( This happens when there is a gap, i.e. all particles have energy $>\epsilon >0$.)  They cannot be applied to QED directly. However, it is well known that inclusive cross sections in QED can be obtained as limits of inclusive cross-sections of the theory with massive photon.  This means that  inclusive cross sections in QED can be obtained from GGreen functions on shell.

\clearpage
\appendix
\setcounter{equation}{0}
\renewcommand{\theequation}{A.\arabic{equation}}
\section{Proper functions and inclusive cross sections in QED}
\label{qed:appendix}

\subsection{QED input for the induction}
\label{qed:induction}
Use metric $g={\rm diag}(1,-1,-1,-1)$ and Feynman gauge for internal photons.
We put $c=1,\hbar=1.$ 

The interaction in the formalism of L-functionals becomes
\begin{equation}
 \cV=-ie_0\sum_f Q_f\int d^4x\,
 \bigl[\bar\psi_{f}^+\gamma^\mu A_{\mu}^+\psi_{f}^+
       -\bar\psi_{f}^-\gamma^\mu A_{\mu}^-\psi_{f}^-\bigr].
 \label{qed:seed}
\end{equation}
Here $Q_f$ is the signed charge in units of $e_0$.  We say that $\cV$ is a seed. It spacifies the starting point of induction.

Physical charged thresholds are denoted by $M_f$ in the same units.

In branch order $(+,-)$ and Fourier convention $e^{-ip(x-y)}$, the two-point free GGreen functions 
(covariances) are $D^{\mu\nu}(p)=-g^{\mu\nu}B(p;0)$ and
$S_f(p)=(\not p+m_f)B(p;m_f)$, where
\begin{equation}
 B(p;m)=
 \begin{pmatrix}
 \dfrac{i}{p^2-m^2+i0} & 2\pi\theta(-p^0)\delta(p^2-m^2)\\[5pt]
 2\pi\theta(p^0)\delta(p^2-m^2) & \dfrac{-i}{p^2-m^2-i0}
 \end{pmatrix}.
 \label{qed:covariances}
\end{equation}
$S_f$ uses the ordered slots $\psi_f(x),\bar\psi_f(y)$. Its $+-$
component includes the sign from graded interchange. Equation
\eqref{eq:rec} uses all four components for GGreen functions and the
single plus branch for ordinary Green functions. The mixed components
have on-shell support, while the diagonal components are boundary
values. Photon detectors project onto the two transverse polarizations.

For the order $(\psi,\bar\psi)$, the covariance and the left--right
Hessian of $\bar\psi M\psi$ have blocks
\begin{equation}
 P_F=\begin{pmatrix}0&S\\-S^{\rm T}&0\end{pmatrix},\qquad
 H_F=\begin{pmatrix}0&-M^{\rm T}\\M&0\end{pmatrix}.
 \label{qed:gradedblocks}
\end{equation}
The transposes include branch and momentum indices. Use the Legendre
convention \eqref{eq:interactionGamma} and the supertrace of Section~2.
In particular, $\frac{1}{2r}\STr[(P_FH_F)^r]=-\frac1r\Tr[(SM)^r]$.
This fixes the sign and multiplicity of a closed charged cycle.

For the cubic seed, $3v=2m+n$ and $L=m-v+1$, with $v$ vertices
and $n$ external arguments. Hence
\begin{equation}
 m=3L+n-3,\qquad \Gamma_{0,0}=\cV,\qquad
 \Gamma_{1,m}=\frac1{2m}\STr[(P\cV^{(2)})^m].
 \label{qed:one}
\end{equation}
The one-loop two-, three-, and four-point kernels are derivatives
of the $m=2,3,4$ terms. To compute a translation-invariant integrand
by \eqref{eq:rec}, fix the external order and Hessian ports. Give the
closing trace a new momentum, route it by successive transfers around
the trace, and retain the lower blocks' loop variables in product order.
Each loop has measure $\int_\ell=\int d^4\ell/(2\pi)^4$. Compare
routings with a common boundary-value and integration prescription.

\subsection{Proper kernels and tree reconstruction}
\label{qed:proper}
Write $N_f(r)=\not r+m_f$, $d_f^+(r)=r^2-m_f^2+i0$, and
$d_0^+(r)=r^2+i0$. For example, the $m=2,3$ cycles on the plus
branch give the raw amputated densities
\begin{align}
 T_{\Sigma,f}(\ell;p)
 &=-e_0^2Q_f^2\frac{\gamma^\alpha N_f(p-\ell)\gamma_\alpha}
                  {d_0^+(\ell)d_f^+(p-\ell)},\label{qed:Sigma}\\
 T_\Pi^{\mu\nu}(\ell;q)
 &=-e_0^2\sum_hQ_h^2
 \frac{\Tr_4[\gamma^\mu N_h(\ell+q)\gamma^\nu N_h(\ell)]}
      {d_h^+(\ell+q)d_h^+(\ell)},\label{qed:Pi}\\
 T_{V,f}^\mu(\ell;p',p)
 &=-e_0^3Q_f^3
 \frac{\gamma^\alpha N_f(p'-\ell)\gamma^\mu N_f(p-\ell)\gamma_\alpha}
      {d_0^+(\ell)d_f^+(p'-\ell)d_f^+(p-\ell)}.
 \label{qed:vertex}
\end{align}
Here $T_{\Sigma,f}=-i\Sigma_f$ and $\Tr_4$ is the Dirac trace. Between
the charged numerators $\not q$ is the difference of the adjacent
free inverses. This gives the routed Ward check
\begin{equation}
 q_\mu T_{V,f}^\mu(\ell;p',p)
 =e_0Q_f[T_{\Sigma,f}(\ell;p)-T_{\Sigma,f}(\ell;p')],
 \qquad q=p'-p.
 \label{qed:ward}
\end{equation}
For \eqref{qed:Pi} the corresponding difference cancels under
translation-compatible integration. Mixed kernels use the same cycle
and the propagators selected by the ordered branch labels.

At $\tau=z=1$, \eqref{eq:interactionGamma} gives
$\varphi=P(J+\Gamma^{(1)}[\varphi])$. Its derivatives reconstruct
connected functions as trees of proper vertices and bridges, with
$C=(1-P\Gamma^{(2)})^{-1}P$ from \eqref{eq:sourceHessian}.
Exponentiating the connected generator gives the full function as a
sum over disjoint unions of these trees.
\begin{center}\small
\begin{tabular}{@{}lp{0.69\linewidth}@{}}
 \toprule
 Contribution & Reconstruction on the branches\\ \midrule
 Born & Join two cubic seeds by a photon or charged bridge.\\
 Virtual & Correct a bridge or cubic vertex, or take the proper
 four-point block. Add the normalized endpoint terms.\\
 Real & Sum the three-vertex trees on each branch and sew over the
 extra photon's momentum and transverse polarization.\\
 \bottomrule
\end{tabular}
\end{center}
The four-charge $m=4$ cycle gives the direct and crossed boxes. Use
crossing and antisymmetrization of the full ordered coefficient for
$e^-\mu^-$, M\o ller and Bhabha scattering, and $e^-e^+\to\mu^-\mu^+$.
The Compton coefficient also gives two-photon annihilation and pair
creation by crossing before forming the interferences.

For instance, in Compton scattering let $1,2$ carry the photon
transfers $k,-k'$ along the charged line and $a,b$ the virtual transfers
$-\ell,\ell$. A word specifies the ordered matrices from the incoming
charged endpoint, with the propagator momentum after each letter given
by the partial sum of transfers. The $m=4$ cycle supplies $a12b$ and
$a21b$, where the virtual photon spans both observed vertices.
Tree reconstruction adds the two bridge insertions $1ab2,2ab1$ and
the four vertex corrections $a1b2,1a2b,a2b1,2a1b$. The charged
endpoint words are accounted for by the two-point reduction.

\subsection{The measured GGreen function}
\label{qed:detectors}
Use the vacuum $\omega_\Omega$ of \eqref{eq:dressing} in the regulated
expansion of \eqref{eq:G}. With incoming creation packet $B_i$, the
two-sided preparation gives the outgoing detector expectation
\begin{equation}
 K_F=\frac{\omega_\Omega(B_i^*O_{F,\rm out}B_i)}
          {\omega_\Omega(B_i^*B_i)}.
 \label{qed:prepared}
\end{equation}
With two incoming and two observed outgoing particles, $K_F$ comes
from the full eight-argument GGreen function, four arguments per
branch. Take its normalized on-shell diagonal on packets with the
observation-time normalization \cite{NP,B,QEDAdiabatic}.

The normalized projection supplies endpoint terms from the slopes of
the proper two-point functions. With $v=p/m_f$ for a charged external
momentum, put $P_v=(1+\not v)/2$ and
\[
 r_f(\ell;v)=\left.\partial_E\,\frac12\Tr_4
       [P_v\Sigma_f(\ell;Ev)]\right|_{E=m_f}.
\]
For a transverse photon of momentum $k$ take the $\zeta$ derivative
of $i\epsilon_\mu^*T_\Pi^{\mu\nu}(\ell;k+\zeta n)\epsilon_\nu$ at
$\zeta=0$, with $n^2=0$, $2k\cdot n=1$, and $\epsilon\cdot n=0$.
The inverse charged propagator at the matched shell has energy slope
$1-\int_\ell r_f$ through one loop. Expanding its inverse square root
within the packet normalization, and doing the same for photons, gives
\begin{equation}
 T_{1,\rm red}(\ell)=T_{\rm conn}^{[1]}(\ell)
                    +\tfrac12 T_0\sum_{\text{external }i}r_i(\ell).
 \label{qed:reduction}
\end{equation}
Use these terms at the common regulator in the inclusive combination.

Write $T=i\mathcal M$ for a contracted branch coefficient of the
regulated evolution and $F_r$ for the detector weight on $r$ final
particles. With ordered phase space $d\Phi_r$, event symmetry factors
$S_r$, and incident flux $\mathcal F$, the cross section through NLO is
$(K_{F,0}+\delta K_F)/\mathcal F$, where
\begin{align}
 K_{F,0}&=\frac1{S_2}\int d\Phi_2\,F_2\,\overline{|\mathcal M_0|^2},
 \nonumber\\
 \delta K_F&=\frac1{S_2}\int d\Phi_2\,F_2\,
   \overline{2\Real\left(\mathcal M_0^*\int_\ell
                      \mathcal M_{1,\rm red}(\ell)\right)}
  +\frac1{S_3}\int d\Phi_3\,F_3\,
           \overline{|\mathcal M_\gamma^{[0]}|^2}.
 \label{qed:NLO}
\end{align}
The overline includes the incoming spin density matrix and the measured
spin or polarization sums. In a tagged-photon observable, use the
symmetry factor for the chosen tagging convention consistently with
$F_3$. The two NLO terms are evaluated with a common soft prescription.

\subsection{Soft cancellation on a common measure}
\label{qed:IR}
Take massive charges, with compact hard support away from thresholds
and forward Coulomb poles. For future-directed charged momenta
$p_i$ and their signed charges $Q_i$ (including antiparticles), define
\begin{equation}
 J_{\rm s}^\mu(k)=\sum_i\eta_i Q_i\frac{p_i^\mu}{p_i\cdot k},
 \qquad \eta_i=\begin{cases}-1&\text{incoming},\\+1&\text{outgoing},\end{cases}
 \qquad k\cdot J_{\rm s}=0.
 \label{qed:softcurrent}
\end{equation}
The external attachments in the real tree give
$e_0\epsilon^*\cdot J_{\rm s}\mathcal M_0+O(1)$ as $k\to0$,
using the on-shell numerator $2p_i^\mu$ and the denominator
$2\eta_i p_i\cdot k$. For four charged legs the virtual vertex
corrections give pairs within a line and the boxes give pairs across
the two lines. For Compton scattering the proper four-point block gives
the pair of distinct legs. Equation \eqref{qed:reduction} supplies the
diagonal pairs. After reducing the virtual photon energy, the leading
densities on $d\Pi_\gamma=d^3\mathbf k/[(2\pi)^3 2|\mathbf k|]$ are
\begin{equation}
 R_{\rm s}=+C_{\rm s},\qquad V_{\rm s}=-C_{\rm s},\qquad
 C_{\rm s}=e_0^2(-J_{\rm s}^2)\overline{|\mathcal M_0|^2}.
 \label{qed:softcancel}
\end{equation}
The linear soft Hamiltonian displaces each mode by
$\beta_\lambda=e_0\epsilon_\lambda^*\cdot J_{\rm s}$ in phase-space
normalization, up to a phase in the evolution. The vacuum and
one-quantum probabilities start as $1-|\beta_\lambda|^2$ and
$|\beta_\lambda|^2$, fixing the two signs in the same state expectation.

For an explicit common prescription, factor
$d\Phi_3(P)=d\Pi_\gamma\,d\Phi_2(P-k)$ and project to the Born
configuration by keeping the two-body direction in the respective
rest frames, identified by rotationless boosts. The phase-space ratio
is $\mathcal J=1+O(\omega)$ with $\omega=|\mathbf k|$ in the chosen
soft-energy frame. In each soft sector subtract $F_2\chi(k) C_{\rm s}$
from the mapped real density $F_3\mathcal J\overline{|\mathcal M_\gamma^{[0]}|^2}$
and add its integral to the virtual term with the same lower bound
$\omega>\kappa$. Here $\chi=1$ near zero and vanishes outside the soft
region. The event or tagging factors in each sector must reproduce
the Born factor in the soft limit.

If $F_3-F_2=O(\omega^a)$ on this projection for some $a>0$, the
leading weighted sum is $(\mathcal J F_3-F_2)C_{\rm s}$.
Since $C_{\rm s}=O(\omega^{-2})$ and
$d\Pi_\gamma\propto\omega\,d\omega\,d\Omega$, the radial bound is
$O(\omega^{\min(a,1)-1})\,d\omega$. The subleading soft terms are
integrable, and the bound is angularly uniform on the stated massive
hard support. Thus the diagonal NLO observable has a $\kappa\downarrow0$
limit with the photon massless throughout.

\subsection{Ultraviolet subtractions and physical inputs}
\label{qed:matching}
Use Ward-compatible subtractions for the UV extension, including the
subtractions of proper subkernels \cite{QEDZimmermann,QEDBonini}.
The Ward identity ties the charged vertex subtraction to the charged
kinetic term. Include the usual QED local action subtractions
$\Delta S_{j,\Lambda}$ in the induction by setting
\begin{equation}
 C_{j,\Lambda}=\frac{i}{\hbar}\bigl(
  \Delta S_{j,\Lambda}[\varphi_+]-\Delta S_{j,\Lambda}[\varphi_-]\bigr),
 \qquad
 \Gamma_{\tau=0,z}=\cV+\sum_{j\geq1}z^j C_{j,\Lambda}.
 \label{qed:localseed}
\end{equation}
Equation \eqref{eq:rec} for $m>0$ is unchanged. The new edge-zero
coefficients are $\Gamma_{j,0}=C_{j,\Lambda}$, and $L$ then counts total
perturbative order including the assigned subtraction grades. Mixed
proper functions inherit lower-subkernel subtractions from these same
seeds. UV cancellation combines edge grades at $\tau=1$.

Equivalently, start with the bare induction and match the physical
inputs in the normalized cross section. Let $g_0=e_0^2$ and write
\begin{equation}
 X_F=\frac{K_F}{\mathcal F}
    =g_0^2 C_F(m_0)+g_0^3 D_F(m_0;\Lambda)+O(g_0^4),
 \qquad M_f=m_{0f}+g_0 A_{f,\Lambda}(m_{0f})+O(g_0^2).
 \label{qed:barematching}
\end{equation}
Here $D_F$ is the inclusive coefficient of \eqref{qed:NLO} after soft
cancellation. With Ward-compatible tensor reduction before the
four-dimensional Euclidean scalar cutoff \cite{QEDCutoff}, the shell
projection of \eqref{qed:Sigma} gives
\begin{equation}
 A_{f,\Lambda}(m)
 =\frac{Q_f^2m}{16\pi^2}
   \left(3\log\frac{\Lambda^2}{m^2}+2\right)+o(1),
 \label{qed:mass}
\end{equation}
Mass matching changes the entire Born coefficient, including flux,
phase space, and any mass-dependent detector definition. Put
$H_{F,\Lambda}=D_F(M;\Lambda)
-\sum_f A_{f,\Lambda}(M_f)\partial_{M_f}C_F(M)$.

Fix the dimensionless physical charge $g=e_{\rm phys}^2/\hbar$ by an
inclusive reference measurement $X_*=g^2C_*(M)$ with $C_*(M)\ne0$.
Eliminating $g_0$ between that measurement and \eqref{qed:barematching}
gives
\begin{equation}
 X_F^{\rm phys}=g^2 C_F+
 g^3\lim_{\Lambda\to\infty}\left[
 H_{F,\Lambda}-\frac{C_F}{C_*}H_{*,\Lambda}\right]+O(g^4).
 \label{qed:physicalanswer}
\end{equation}
The Ward identity cancels the charged kinetic and vertex logarithms
in the normalized reduction. The photon cycle \eqref{qed:Pi} enters
through the photon bridge or the two photon endpoint slopes and gives
\[
 H_{F,\Lambda}=-2B_{\rm QED}C_F\log(\Lambda/\mu)+O(1),\qquad
 B_{\rm QED}=\frac1{6\pi^2}\sum_h Q_h^2.
\]
The logarithm cancels in \eqref{qed:physicalanswer}. With
\eqref{qed:localseed} the same measurements fix the finite parts of
the locally subtracted coefficients.

\clearpage
\setcounter{equation}{0}
\renewcommand{\theequation}{B.\arabic{equation}}
\section{Executable induction for proper functions}
\label{code:appendix}

This appendix gives a complete Python implementation of
\eqref{eq:rec}, including its fermionic version. The input is a
field-independent covariance $P$ and an even polynomial interaction
$\cV$. The output is the interaction proper functional
$\Gamma_{L,m}$ and any of its ordered field derivatives at zero
field. The free quadratic term is removed, exactly as in
\eqref{eq:interactionGamma}. Both ordinary proper Green functions
and proper GGreen functions use this program: in the second case
the field indices include the two contour branches and $P$ is the
complete doubled covariance.

The implementation works with finitely many modes, as in Section~2.
Thus it also applies directly after a finite-volume and finite-mode
regularization. It handles arbitrary polynomial degree, mixed bosonic
and fermionic fields, and degenerate covariances. All coefficients are
exact rational numbers or polynomials in commuting symbolic parameters.
Python~3.10 or later is required; there are no external Python
dependencies.

\subsection{A convolution form of the same induction}

Writing $t=\tau$ in the code, introduce the matrix series
\begin{equation}
 A(z,t)=P\Gamma^{(2)}(z,t),\qquad
 R(z,t)=(1-tA(z,t))^{-1}
        =\sum_{r,s\geq0}z^rt^sR_{r,s}.
 \label{code:resolvent}
\end{equation}
The identity $R=1+tAR$ gives
\begin{align}
 R_{0,0}&=1,\qquad R_{r,0}=0\quad(r>0),\nonumber\\
 R_{r,s}&=\sum_{\ell=0}^{r}\sum_{e=0}^{s-1}
       \bigl(P\Gamma_{\ell,e}^{(2)}\bigr)
       R_{r-\ell,s-1-e}\qquad(s>0).
 \label{code:convolution}
\end{align}
Consequently the recurrence implemented below is
\begin{equation}
 \Gamma_{L,m}=\frac{1}{2m}\STr R_{L-1,m}
 \quad(L,m>0),\qquad
 \Gamma_{0,0}=\cV,
 \label{code:gamma}
\end{equation}
with the remaining $L=0$ or $m=0$ coefficients zero.
Expanding $R$ as a geometric series recovers exactly
\eqref{eq:rec}; no new approximation has been introduced.
The calculation of $R_{L-1,m}$ uses proper functions of strictly
lower loop order, while its further calls to $R$ have smaller
edge index. These two facts prove termination.

Every calculated coefficient and Hessian product is cached.
For bounds $L_{\max},m_{\max}$ there are
$O(L_{\max}m_{\max})$ cached matrix coefficients and
$O(L_{\max}^2m_{\max}^2)$ matrix-convolution summands.
These counts treat differentiation and polynomial matrix operations
as operations on whole objects. The numbers of polynomial monomials,
field components, and elementary arithmetic operations can grow
substantially; these counts are not a polynomial bound on the full
QED calculation.

\subsection{Input conventions and extraction of proper kernels}

\texttt{Ring} records the bosonic fields, the ordered Grassmann fields,
and the commuting parameters. Its field order is always bosons first,
then fermions. The order of the Grassmann fields fixes the canonical
order of their monomials. For example,
\begin{lstlisting}[style=properpython]
r = Ring(bosons=("a",), fermions=("psi", "bar"),
         parameters=("g", "d", "s"))
a, psi, bar, g, d, s = (r.symbol(name) for name in
                       ("a", "psi", "bar", "g", "d", "s"))
P = [[d, 0, 0], [0, 0, s], [0, -s, 0]]
c = Proper(r, P, g * bar * a * psi)
print(c.gamma(1, 2))
print(c.vertex(1, 2, ("psi", "bar")))
print(c.calculate(max_loop=2, max_edge=3))
\end{lstlisting}
The first output represents
$g^2ds\bar\psi\psi-\tfrac12g^2s^2a^2$, in canonical odd order.
The second output is $-g^2ds$, because the supplied leg order means
first a left derivative in $\psi$, then one in $\bar\psi$.
This explicitly specifies the external fermionic ordering.

For an even polynomial $F$, the Hessian used inside the induction is
\begin{equation}
 H_{AB}(F)=\partial^R_B\bigl(\partial^L_A F\bigr).
 \label{code:hessian}
\end{equation}
In particular it gives exactly the blocks
\eqref{qed:gradedblocks} for $F=\bar\psi M\psi$.
Matrix multiplication keeps coefficient order and the supertrace
multiplies the $A$th diagonal entry by $(-1)^{|A|}$.
The code checks that the input covariance is even, independent of
the fields, and graded symmetric. Its entries have commuting
coefficients; mixed even--odd entries therefore vanish.

\texttt{vertex(L,m,legs)} applies successive left derivatives in the
listed order and then sets every field to zero. It returns the
derivative itself, including its factorials, rather than the
coefficient of a monomial. Bosonic and fermionic vacuum terms are
retained during induction. There is no intermediate truncation in
external field degree: higher-degree lower-loop terms can contribute
to lower-degree higher-loop kernels.

Each propagator entry and each interaction coefficient can be an
independent commuting parameter. In particular a denominator or a
complex kernel value can be represented by a symbol for the entire
entry. After extracting a vertex, \texttt{evaluate(values)} substitutes
its parameter values; these can also be symbolic expressions supplied
by an external computer algebra system. All phases, branch signs,
symmetry factors in the seed, and factors of $\hbar$ belong to the
input. The formal loop grade is not identified with $\hbar$.
For example, for $\cV=gx^4/4!$ and $P=p$,
\begin{lstlisting}[style=properpython]
r = Ring(bosons=("x",), parameters=("g", "p"))
x, g, p = (r.symbol(name) for name in ("x", "g", "p"))
c = Proper(r, [[p]], g * x**4 / 24)
k = c.vertex(2, 3, ("x", "x"))  # 5*g**2*p**3/12
print(k.evaluate({"g": Q(1, 10), "p": Q(2)}))  # 1/30
\end{lstlisting}

For QED one supplies a separate mode coordinate for every photon
component, charged spinor component, and branch label. The charged
covariance has the ordered blocks $(0,S;-S^{\rm T},0)$ of
\eqref{qed:gradedblocks}; the interaction is the ordered seed
\eqref{qed:seed}. Ordinary Green functions use the single-branch
input. The matrix sums below are the finite-mode contractions; no
momentum integral is evaluated by the program. A continuum calculation
additionally requires the momentum routing, measures, and common
boundary-value and regulator prescriptions discussed in Appendix~A.
Inclusive observables are reconstructed from these proper kernels
as described there.

\subsection{Complete source code}

Save the following listing as \texttt{proper\_functions.py}.
It includes exact self-tests, which run with the command
\begin{center}\texttt{python3 proper\_functions.py}\end{center}
Importing the file exposes the classes without running the tests.

\begin{lstlisting}[style=properpython]
"""Proper functions by the loop/edge induction of QFTQED.tex.

Python 3.10+, standard library only. Run this file for exact checks.
Finite-mode bosonic/Grassmann polynomials; no momentum integration.
Propagator entries may be commuting symbols for arbitrary kernels.
"""

from fractions import Fraction as Q
from functools import lru_cache


class Ring:
    """Bosonic fields, ordered odd fields, and even parameters."""

    def __init__(self, bosons=(), fermions=(), parameters=()):
        self.bosons = tuple(bosons)
        self.fermions = tuple(fermions)
        self.parameters = tuple(parameters)
        self.even = self.bosons + self.parameters
        self.fields = self.bosons + self.fermions
        names = self.even + self.fermions
        if len(set(names)) != len(names):
            raise ValueError("Variable names must be distinct")
        self.parity = (0,) * len(bosons) + (1,) * len(fermions)
        self.unit_key = ((0,) * len(self.even), 0)
        self.zero = Poly(self, {})
        self.one = Poly(self, {self.unit_key: Q(1)})

    def coerce(self, value):
        if isinstance(value, Poly):
            if value.ring is not self:
                raise ValueError("Different polynomial rings")
            return value
        return Poly(self, {self.unit_key: Q(value)})

    def symbol(self, name):
        powers = [0] * len(self.even)
        mask = 0
        if name in self.even:
            powers[self.even.index(name)] = 1
        else:
            mask = 1 << self.fermions.index(name)
        return Poly(self, {(tuple(powers), mask): Q(1)})


class Poly:
    """Sparse polynomial; odd monomials follow Ring.fermions order.

    Treat instances as immutable. Coefficients are exact rationals;
    symbolic propagators and couplings are even parameters.
    """

    def __init__(self, ring, terms):
        self.ring = ring
        self.terms = {k: Q(v) for k, v in terms.items() if v}

    def __bool__(self):
        return bool(self.terms)

    def __eq__(self, other):
        other = self.ring.coerce(other)
        return self.terms == other.terms

    def __add__(self, other):
        other = self.ring.coerce(other)
        terms = self.terms.copy()
        for key, value in other.terms.items():
            terms[key] = terms.get(key, Q(0)) + value
        return Poly(self.ring, terms)

    __radd__ = __add__

    def __neg__(self):
        return Poly(self.ring, {k: -v for k, v in self.terms.items()})

    def __sub__(self, other):
        return self + (-self.ring.coerce(other))

    def __rsub__(self, other):
        return self.ring.coerce(other) - self

    def __mul__(self, other):
        other = self.ring.coerce(other)
        terms = {}
        for (pa, ma), ca in self.terms.items():
            for (pb, mb), cb in other.terms.items():
                if ma & mb:
                    continue  # Repeated odd variable: square is zero.
                swaps, remaining = 0, ma
                while remaining:
                    bit = remaining & -remaining
                    swaps += (mb & (bit - 1)).bit_count()
                    remaining ^= bit
                key = (tuple(a + b for a, b in zip(pa, pb)), ma | mb)
                value = ca * cb * (-1 if swaps % 2 else 1)
                terms[key] = terms.get(key, Q(0)) + value
        return Poly(self.ring, terms)

    def __rmul__(self, other):
        return self.ring.coerce(other) * self

    def __truediv__(self, scalar):
        scalar = Q(scalar)
        if not scalar:
            raise ZeroDivisionError
        return Poly(self.ring, {k: v / scalar
                                for k, v in self.terms.items()})

    def __pow__(self, power):
        if not isinstance(power, int) or power < 0:
            raise ValueError("Only nonnegative integer powers")
        result, base = self.ring.one, self
        while power:
            if power % 2:
                result = result * base
            base = base * base
            power //= 2
        return result

    def diff(self, name, right=False):
        """Left derivative by default; right=True gives right one."""
        ring, terms = self.ring, {}
        if name in ring.even:
            i = ring.even.index(name)
            for (powers, mask), value in self.terms.items():
                if powers[i]:
                    new = list(powers)
                    new[i] -= 1
                    terms[(tuple(new), mask)] = value * powers[i]
        else:
            i = ring.fermions.index(name)
            bit = 1 << i
            for (powers, mask), value in self.terms.items():
                if mask & bit:
                    count = ((mask >> (i + 1)).bit_count() if right
                             else (mask & (bit - 1)).bit_count())
                    terms[(powers, mask ^ bit)] = value * (-1)**count
        return Poly(ring, terms)

    def at_zero_fields(self):
        nb = len(self.ring.bosons)
        return Poly(self.ring, {k: v for k, v in self.terms.items()
                               if k[1] == 0 and not any(k[0][:nb])})

    def vertex(self, legs):
        """Successive LEFT derivatives in the supplied leg order,
        evaluated at zero fields; includes factorials automatically.
        """
        result = self
        for name in legs:
            if name not in self.ring.fields:
                raise ValueError("A vertex leg must be a field")
            result = result.diff(name)
        return result.at_zero_fields()

    def evaluate(self, values):
        """Evaluate a field-independent result after supplying every
        used parameter. Values can be numeric or external CAS objects.
        """
        if self != self.at_zero_fields():
            raise ValueError("Extract a vertex before evaluating")
        total = 0
        for (powers, _), coefficient in self.terms.items():
            term = coefficient
            for name, power in zip(self.ring.even, powers):
                if power:
                    term = term * values[name]**power
            total = total + term
        return total

    def __repr__(self):
        if not self:
            return "0"
        pieces = []
        for (powers, mask), coefficient in sorted(self.terms.items()):
            factors = [str(coefficient)]
            for name, power in zip(self.ring.even, powers):
                if power:
                    factors.append(name if power == 1
                                   else f"{name}**{power}")
            factors += [name for i, name in enumerate(self.ring.fermions)
                        if mask & (1 << i)]
            pieces.append("*".join(factors))
        return " + ".join(pieces).replace("+ -", "- ")


def zeros(ring):
    n = len(ring.fields)
    return tuple(tuple(ring.zero for _ in range(n)) for _ in range(n))


def identity(ring):
    n = len(ring.fields)
    return tuple(tuple(ring.coerce(int(i == j)) for j in range(n))
                 for i in range(n))


def madd(a, b):
    return tuple(tuple(x + y for x, y in zip(ar, br))
                 for ar, br in zip(a, b))


def mmul(a, b):
    # Preserve coefficient order: Grassmann entries do not commute.
    n = len(a)
    return tuple(tuple(sum((a[i][k] * b[k][j] for k in range(n)),
                           a[0][0].ring.zero)
                       for j in range(n)) for i in range(n))


def hessian(f):
    # H_ab = right_d_b(left_d_a F), as in Appendix A.
    return tuple(tuple(f.diff(a).diff(b, right=True)
                       for b in f.ring.fields) for a in f.ring.fields)


def strace(a, ring):
    return sum(((-1)**p * a[i][i]
                for i, p in enumerate(ring.parity)), ring.zero)


class Proper:
    """Cached induction for Gamma[L,m], with the free term removed.

    P is a field-independent, even, graded-symmetric covariance.
    V is the EVEN interaction polynomial, including all phases/signs.
    Coordinates are ordered as ring.fields. No graph enumeration.
    """

    def __init__(self, ring, propagator, interaction):
        self.ring = ring
        n = len(ring.fields)
        if not n or len(propagator) != n:
            raise ValueError("Propagator dimension must match fields")
        if any(len(row) != n for row in propagator):
            raise ValueError("Propagator must be square")
        self.P = tuple(tuple(ring.coerce(x) for x in row)
                       for row in propagator)
        self.V = ring.coerce(interaction)
        if any(mask.bit_count() % 2 for _, mask in self.V.terms):
            raise ValueError("Interaction must be even")
        for i, pi in enumerate(ring.parity):
            for j, pj in enumerate(ring.parity):
                entry = self.P[i][j]
                if entry != entry.at_zero_fields():
                    raise ValueError("P must be field independent")
                if pi != pj and entry:
                    raise ValueError("No odd external coefficients")
                if entry != (-1)**(pi * pj) * self.P[j][i]:
                    raise ValueError("P must be graded symmetric")
        self.Z = zeros(ring)
        self.I = identity(ring)

    @lru_cache(maxsize=None)
    def A(self, loop, edge):
        return mmul(self.P, hessian(self.gamma(loop, edge)))

    @lru_cache(maxsize=None)
    def R(self, loop, edge):
        """Coefficient [z**loop t**edge] of (I - t*A)**(-1)."""
        if edge == 0:
            return self.I if loop == 0 else self.Z
        answer = self.Z
        for ell in range(loop + 1):
            for e in range(edge):
                a = self.A(ell, e)
                if not any(any(row) for row in a):
                    continue
                b = self.R(loop - ell, edge - 1 - e)
                answer = madd(answer, mmul(a, b))
        return answer

    @lru_cache(maxsize=None)
    def gamma(self, loop, edge):
        """The complete polynomial Gamma_{loop,edge}."""
        if not isinstance(loop, int) or not isinstance(edge, int):
            raise ValueError("Grades must be integers")
        if loop < 0 or edge < 0:
            raise ValueError("Grades must be nonnegative")
        if loop == 0 and edge == 0:
            return self.V
        if loop == 0 or edge == 0:
            return self.ring.zero
        return strace(self.R(loop - 1, edge), self.ring) / (2 * edge)

    def calculate(self, max_loop, max_edge):
        if max_loop < 0 or max_edge < 0:
            raise ValueError("Bounds must be nonnegative")
        return {(ell, m): self.gamma(ell, m)
                for ell in range(max_loop + 1)
                for m in range(max_edge + 1)}

    def vertex(self, loop, edge, legs):
        return self.gamma(loop, edge).vertex(legs)


def self_test():
    r = Ring(bosons=("x",), parameters=("g", "p"))
    x, g, p = (r.symbol(s) for s in ("x", "g", "p"))
    c = Proper(r, [[p]], g * x**4 / 24)
    assert c.gamma(1, 1) == g * p * x**2 / 4
    assert c.gamma(1, 2) == g**2 * p**2 * x**4 / 16
    assert c.gamma(2, 2) == g * p**2 / 8
    assert c.gamma(2, 3) == 5 * g**2 * p**3 * x**2 / 24
    assert c.vertex(2, 3, ("x", "x")) == 5 * g**2 * p**3 / 12
    # Exact quadratic resummation has no terms above one loop.
    q = Proper(r, [[p]], g * x**2 / 2)
    for m in range(1, 7):
        assert q.gamma(1, m) == (p * g)**m / (2 * m)
        assert q.gamma(2, m) == 0
    # Minimal Yukawa example: check both odd algebra and loop signs.
    r = Ring(bosons=("a",), fermions=("psi", "bar"),
             parameters=("g", "d", "s"))
    a, psi, bar, g, d, s = (r.symbol(name) for name in
                           ("a", "psi", "bar", "g", "d", "s"))
    assert psi * bar == -bar * psi
    assert psi**2 == 0
    assert (psi * bar).diff("bar") == -psi
    assert (psi * bar).diff("bar", right=True) == psi
    c = Proper(r, [[d, 0, 0], [0, 0, s], [0, -s, 0]],
               g * bar * a * psi)
    assert c.gamma(1, 1) == -g * s * a
    assert c.gamma(1, 2) == (g**2 * d * s * bar * psi
                            - g**2 * s**2 * a**2 / 2)
    assert c.gamma(2, 3) == -g**2 * d * s**2 / 2
    # Off-diagonal bosonic covariance, including a degenerate one.
    r = Ring(bosons=("x", "y"), parameters=("g",))
    x, y, g = (r.symbol(name) for name in ("x", "y", "g"))
    for matrix, expected in (([[2, 1], [1, 3]], 2),
                             ([[1, 1], [1, 1]], Q(3, 4))):
        c = Proper(r, matrix, g * x**2 * y**2 / 4)
        assert c.gamma(2, 2) == expected * g
    print("Exact self-tests passed.")


if __name__ == "__main__":
    self_test()
\end{lstlisting}

\subsection{Checks of coefficients and signs}

The included self-tests give, for $\cV=gx^4/24$ and $P=p$,
\begin{align}
 \Gamma_{1,1}&=\frac{gp}{4}x^2,&
 \Gamma_{1,2}&=\frac{g^2p^2}{16}x^4,\nonumber\\
 \Gamma_{2,2}&=\frac{gp^2}{8},&
 \Gamma_{2,3}&=\frac{5g^2p^3}{24}x^2.
 \label{code:quarticchecks}
\end{align}
For the mixed seed $g\bar\psi a\psi$ above they give
\begin{equation}
 \Gamma_{1,1}=-gsa,\qquad
 \Gamma_{1,2}=g^2ds\bar\psi\psi-\frac12g^2s^2a^2,
 \qquad \Gamma_{2,3}=-\frac12g^2ds^2.
 \label{code:yukawachecks}
\end{equation}
The quadratic-seed tests verify that only the one-loop coefficients
remain. Off-diagonal bosonic contractions and a degenerate covariance
are also checked.

The accompanying archive includes \texttt{verify\_induction.py}.
This script performs the further independent checks listed below.
For the scalar checks the Gaussian moments were calculated first, followed
by the formal logarithm, inversion of $\varphi=W'(J)$, and the
Legendre transform $W-J\varphi+\varphi^2/2$. This route uses neither
the Hessian flow nor the induction. At unit scalar covariance and
$\tau=1$ the comparisons were:
\begin{center}\small
\begin{tabular}{@{}>{\raggedright\arraybackslash}p{0.37\linewidth}
                    >{\raggedright\arraybackslash}p{0.55\linewidth}@{}}
\toprule
Input or identity & Exact comparison performed\\\midrule
$gx^3/3!$ & Gaussian/logarithm/Legendre result through $g^4$,
all loop grades through $L=3$.\\
$gx^4/4!$ & The same independent construction through $g^3$,
all loop grades through $L=4$.\\
$g\bar\psi a\psi$ & Bosonic sector through $g^4$, using the exact
Berezin integral $G(J)=e^{J^2/(2z)}(1-gJ)$.\\
$\bar\psi M\psi$, two charged modes &
$\Gamma_{1,m}=-\Tr[(SM)^m]/m$ for $m=1,\ldots,6$;
higher-loop coefficients vanish.\\
\eqref{eq:rec} & Explicit ordered-composition sums agree with the
cached convolution for a mixed quartic/Yukawa seed,
$L\leq3$, $m\leq4$.\\
\bottomrule
\end{tabular}
\end{center}
For the homogeneous scalar seeds, coupling order and loop grade fix
the internal-edge number, so these independent scalar comparisons
also test the edge grading. All the displayed checks passed with
exact rational arithmetic. They validate the finite-mode algebra
and recurrence; the analytic continuum and inclusive limits require
the additional estimates and prescriptions stated in the paper.

\end{document}